\documentclass[prl,aps,twocolumn,floatfix]{revtex4}

\usepackage{amssymb,graphicx}
\usepackage{epsfig}
\usepackage[usenames]{color}

\def\apjl{Ap. J. Letters}
\def\mnras{MNRAS}
\def\aap{A\&A}

\newcommand{\had}{{\sc had}}

\begin{document}

\title{Binary black holes' effects on electromagnetic fields}

\author{Carlos Palenzuela${}^1$, Matthew Anderson${}^{2,3}$, 
Luis Lehner${}^{2,4,5}$, Steven L. Liebling${}^6$, David Neilsen${}^3$}

\affiliation{
${}^1$Max-Planck-Institut f\" ur Gravitationsphysik, Albert-Einstein-Institut, Golm, Germany\\
${}^2$Department of Physics and Astronomy, Louisiana State University, Baton Rouge, LA 70803, \\
${}^3$Department of Physics and Astronomy, Brigham Young University, Provo, UT 84602, \\
${}^4$Perimeter Institute for Theoretical Physics, Waterloo, Ontario, N2L 2Y5, Canada,\\ 
${}^5$Department of Physics, University of Guelph, Guelph, Ontario N1G 2W1, Canada, \\
${}^6$Department of Physics, Long Island University--C.W. Post Campus, Brookville, NY 11548
}

\date{\today}

%%%%%%%%%%%%%%%%%%%%%%%%%%%%%%%%%%%%%%%%%%%%%%%%%%%%%%%%%%%%%%%%%%%%
%
%   A B S T R A C T
%
%%%%%%%%%%%%%%%%%%%%%%%%%%%%%%%%%%%%%%%%%%%%%%%%%%%%%%%%%%%%%%%%%%%%
\begin{abstract}
In addition to producing gravitational waves (GW), the dynamics
of a binary black hole system could induce emission of electromagnetic (EM) radiation
by affecting the behavior of plasmas and electromagnetic fields
in their vicinity. We here study how the electromagnetic fields are affected 
by a pair of  orbiting black holes through the merger. 
In particular, we show how the binary's dynamics induce
a  variability in possible electromagnetically induced emissions as well as a
possible enhancement of electromagnetic fields during the late-merge and merger epochs. 
These time dependent features will likely leave their imprint in processes
generating detectable emissions and can be exploited in the
detection of electromagnetic counterparts of gravitational waves. 
\end{abstract}

\maketitle

%%%%%%%%%%%%%%%%%%%%%%%%%%%%%%%%%%%%%%%%%%%%%%%%%%%%%%%%%%%%%%%%%%%%
% PLACE FOR COMMENTS
%%%%%%%%%%%%%%%%%%%%%%%%%%%%%%%%%%%%%%%%%%%%%%%%%%%%%%%%%%%%%%%%%%%%
%%%%%%%%%%%%%%%%%%%%%%%%%%%%%%%%%%%%%%%%%%%%%%%%%%%%%%%%%%%%%%%%%%%%
%
%   I N T R O D U C T I O N
%
%%%%%%%%%%%%%%%%%%%%%%%%%%%%%%%%%%%%%%%%%%%%%%%%%%%%%%%%%%%%%%%%%%%%
%\section{Introduction}
\noindent{\bf{\em Introduction:}}
As the era approaches when the detection and analysis of gravitational waves  joins electromagnetic observations,
studying a number of astrophysical systems in both wavebands 
becomes an exciting prospect.
Indeed, most astrophysical systems which produce gravitational waves are very likely also electromagnetically bright
(see e.g.~\cite{Sylvestre:2003vc,Stubbs:2007mk}), with the exception of systems composed solely of black holes.\\
\indent
One example of such  a system is a black hole  surrounded by an accretion disk.
This scenario has been used to explain a vast range of spectacular
phenomena, from AGNs and quasars to gamma ray bursts, all of which are prodigious emitters of
electromagnetic energy. Strong emission from these systems is understood as the
result of radiative processes within jets powered by the extraction of rotational 
and binding energy.\\
\indent
The pioneering work of Penrose~\cite{Penrose:1969pc} and Blandford and
Znajeck~\cite{Blandford:1977ds}, together with  a large body of subsequent work, has provided a basic understanding of
possible mechanisms to explain highly energetic emissions from single black hole systems
interacting with surrounding plasmas.  The interaction of electromagnetic
field lines with the strong gravitational field of a rotating black hole is the fundamental component of these mechanisms
to explain the  acceleration of
particles that traverse the black hole's ergosphere. The scenario of a
pseudo-stationary, single black hole  interacting with an accretion disk is reasonably well understood, and 
this system is
employed to explain energetic phenomena such as gamma ray bursts, AGNs, quasars, blazars, etc. However,
a highly dynamical stage may occur
prior to such a  pseudo-stationary regime which could
give rise to strong emissions.\\
\indent
A related system, studied here, 
involves the collision of
supermassive black holes, whose gravitational radiation would be detectable by the Laser Interferometric
Space Antenna (LISA).
To date, strong
observational evidence indicates that massive black holes
exist in the centers of most galaxies, and that galaxies have undergone mergers in their past. In galactic collisions, the
initial black holes eventually merge as their orbits shrink through a variety of mechanisms. As
discussed in~\cite{1980Natur.287..307B,Milosavljevic:2004cg,Haiman:2009te}, a circumbinary disk is formed as the
binary hollows out the surrounding gas
% Can we leave out this?
%though some gas might remain within a region in between the black holes
(see, e.g.~\cite{1996ApJ...467L..77A,2008ApJ...672...83M,2009MNRAS.393.1423C}).
As the distance between the black holes decreases, the dynamics of the binary become governed by the emission of gravitational waves
and disconnected from the properties of the disk~\cite{2002ApJ...567L...9A,2003MNRAS.340..411L}.\\
\indent
While the late time dynamics of the system can
be understood in terms of the pseudo-stationary picture mentioned above, an interesting intermediate regime 
would include magnetic fields, anchored at the circumbinary disk, being influenced by the orbiting behavior
of the two black holes.
In particular, both precursor emission and enhancement of electromagnetic energy 
could be possible as the electromagnetic fields interact with the gas being stirred and compressed by the black holes' shrinking
orbit. Consequently, the late pre-merger and merger dynamics of a supermassive binary
black hole system  could have a strong influence on the possible interaction between the remaining gas and the
electromagnetic fields and even seed several emission mechanisms more strongly. 
In this work we study such
a scenario by considering the Einstein-Maxwell system in a setup that describes a pair of black holes close to the
merger epoch, and we examine the electromagnetic field behavior and compare to the single black hole case. While we do
not consider a plasma in our current study, our analysis helps to understand the possible behavior in its
presence and lay the foundations to future work in this direction.\\
%Our work is organized as follows, 
%in Section~\secapproach, we briefly review our formulation of the problem and numerical approach.  
%Section III describes our initial configuration. Section IV discusses the solutions obtained
%for the electromagnetic field anchored in a disk interacting with both single and binary black holes 
%and highlight main features which could induce emission with particular patterns.
%Section V offers some final considerations.   
%
%%%%%%%%%%%%%%%%%%%%%%%%%%%%%%%%%%%%%%%%%%%%%%%%%%%%%%%%%%%%%%%%%%%%
%
%   S E C T I O N
%
%%%%%%%%%%%%%%%%%%%%%%%%%%%%%%%%%%%%%%%%%%%%%%%%%%%%%%%%%%%%%%%%%%%%
%\section{Formulation and numerical approach}
\noindent{\bf{\em Overview of the numerical approach:}} 
We solve the coupled Einstein-Maxwell system to model the black hole merger interacting with
an externally sourced magnetic field. 
The Einstein equations
are written in the Generalized Harmonic formalism
described in~\cite{Palenzuela:2006wp}. The Maxwell equations are written directly 
in terms of the electric and magnetic fields, as in~\cite{2007MNRAS.382..995K}. 
The electromagnetic fields have astrophysically motivated
magnitudes, and thus their energy is several orders of magnitude smaller than the gravitational field energy. Hence the dynamical
effects of the electromagnetic fields on the geometry are negligible. This fact is corroborated
by simulations not displaying any difference in the black holes dynamics with an initial magnetic field 
an order of magnitude larger and even zero, than the one employed here. 

In both systems of equations,
the constraints are kept under control via damping mechanisms: constraint damping
for the Einstein equations~\cite{Gundlach:2005eh} and divergence cleaning for the Maxwell equations
(similar to that defined in~\cite{Dedner2002} for ideal MHD).
A combination of Sommerfeld and constraint preserving boundary conditions are applied
at the outer boundary~\cite{RLS07} for both systems. 
The incoming modes of the electromagnetic fields are defined via Dirichlet 
(maximally dissipative) conditions induced from the physical picture
of a circumbinary disk present beyond the computational domain.

We use the \had\ computational infrastructure that provides distributed Berger-Oliger
style Adaptive Mesh Refinement (AMR)~\cite{had_webpage,Liebling} with full sub-cycling
in time, together with a novel treatment of artificial boundaries~\cite{Lehner:2005vc}.
The refinement regions are determined using truncation error estimation via a shadow
hierarchy. 
A fourth order spatial discretization satisfying a summation by parts rule
together with a third order Runge-Kutta scheme for the time integration are used to help
ensure stability
of the numerical implementation~\cite{binaryNS}.

%%%%%%%%%%%%%%%%%%%%%%%%%%%%%%%%%%%%%%%%%%%%%%%%%%%%%%%%%%%%%%%%%%%%
%
%   S E C T I O N
%
%%%%%%%%%%%%%%%%%%%%%%%%%%%%%%%%%%%%%%%%%%%%%%%%%%%%%%%%%%%%%%%%%%%%
%\section{Binary neutron stars}

\noindent{\bf{\em Overview of the physical set-up:}} 
To explore the effects of the merger dynamics on the electromagnetic field,
we compare certain cases of a single spinning black hole with cases of merging holes.
In all cases, the orbital plane of the evolution (or equatorial plane for the single BH)
is assumed to be aligned with that of the circumbinary disk. The magnetic field
is defined as anchored in the disk; hence, its associated magnetic dipole is aligned
with the orbital and spin angular momentum.

As we are primarily interested in the dynamics close to the merger, we adopt initial data 
such that it takes place after about one orbit.
These data describe quasi-equilibrium, equal-mass, non-spinning black holes
constructed by the publicly available {\sc lorene} code~\cite{lorene_webpage}.
The black holes have masses
given by $M_s = M/2$, and are initially separated by $\approx 6 M$, lying beyond
the approximate inner most stable circular orbit (ISCO)~\cite{Buonanno:2007sv}. 
The initial magnetic field is poloidal and constructed from the electromagnetic potential
produced by a circular loop, whose radius is assumed to be larger
than the domain of interest~\cite{Jackson1975}. We assume the disk lies
at $\simeq 10^3M$ and so $B^i \simeq B_o \hat z$.
The electric field is initially zero throughout and
the magnetic field strength adopted  
is $B_o = 10^{4} (M/10^8 M_{\odot})$~G, which is within values inferred in 
relevant astrophysical systems (see e.g.~\cite{2008A&A...477....1M}). 

We adopt a cubical domain given by
$[-125 \,M, 125 \,M]^3$ and employ an AMR configuration with 
$7$ levels of refinement that adjust themselves dynamically to ensure that
the solution's error is below a pre-determined threshold using
a shadow hierarchy. Residuals of the various constraints and convergence of
the physical fields with resolution are checked as tests of the code.

\begin{figure}[h]
\begin{center}
\epsfig{file=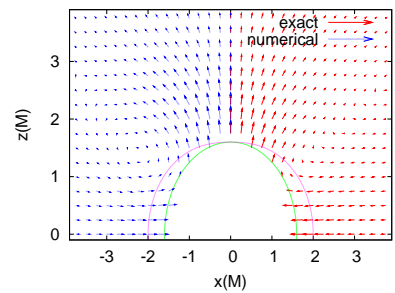,height=4.cm,width=5.5cm}
\caption{The electric field 
for $y=0,z\geq 0$  at $t=200M$ for a
single black hole, together with the apparent horizon~(green) and the ergosphere~(magenta).
The region $x<0$ display the numerical solution,
while Wald's exact solution in $x>0$.}
\label{fig:single_field}
\end{center}
\end{figure}

\noindent{\bf{\em Results:}}
For concreteness we present the results of a single black hole with spin 
parameter $a=0.7/M_s$. We adopt this value, which is slightly larger than the
final spin expected for a merged black hole~\footnote{ 
Calculated directly via simulations 
(for recent efforts in simulations and data analysis 
see e.g.~\cite{Aylott:2009ya} and references cited therein) 
or estimated by simple arguments as in~\cite{Buonanno:2007sv}.}
for comparison purposes.
The evolution shows an initial transient, where
the EM fields adapt to the geometry of the black hole spacetime,
giving rise to an electric field and a deformation of the magnetic field  (see also~\cite{KiLaKu75,Komissarov:2007rc}). 
After $t \simeq 40-60M$
the solution is clearly seen to evolve towards a quasi-stationary state determined by 
Wald's solution~\cite{1974PhRvD..10.1680W} for a Kerr black hole immersed in a uniform
magnetic field which is aligned with its spin. Fig.~\ref{fig:single_field} presents
%%both 
the electric 
%and magnetic 
field obtained at $t=200M$ in
the plane $y=0$. For comparison purposes, the corresponding field from Wald's exact solution is shown for
$x>0$. The apparent agreement along with a careful examination of the asymptotic
solution indicates that,
for a black hole immersed within an almost uniform magnetic field aligned with its spin,
the final state is  Wald's solution~\cite{KiLaKu75}.

\vspace{-0.15in} 

%\begin{figure}[h]
%\begin{center}
%\epsfig{file=figures/single_energyEMF1_t200.eps,height=4.2cm,width=4.2cm} 
%\epsfig{file=figures/single_energyEMF3_t200.eps,height=4.2cm,width=4.2cm} 
%\caption{Snapshots of the electromagnetic energy for a single spinning
%black hole at $t/M\simeq 200$. On the left column is plotted the plane
% $x=0$ while the $z=0$ plane is on the right.} 
%\label{fig:binary_B}
%\end{center}
%\end{figure}

\begin{figure}[h]
\begin{center}
\epsfig{file=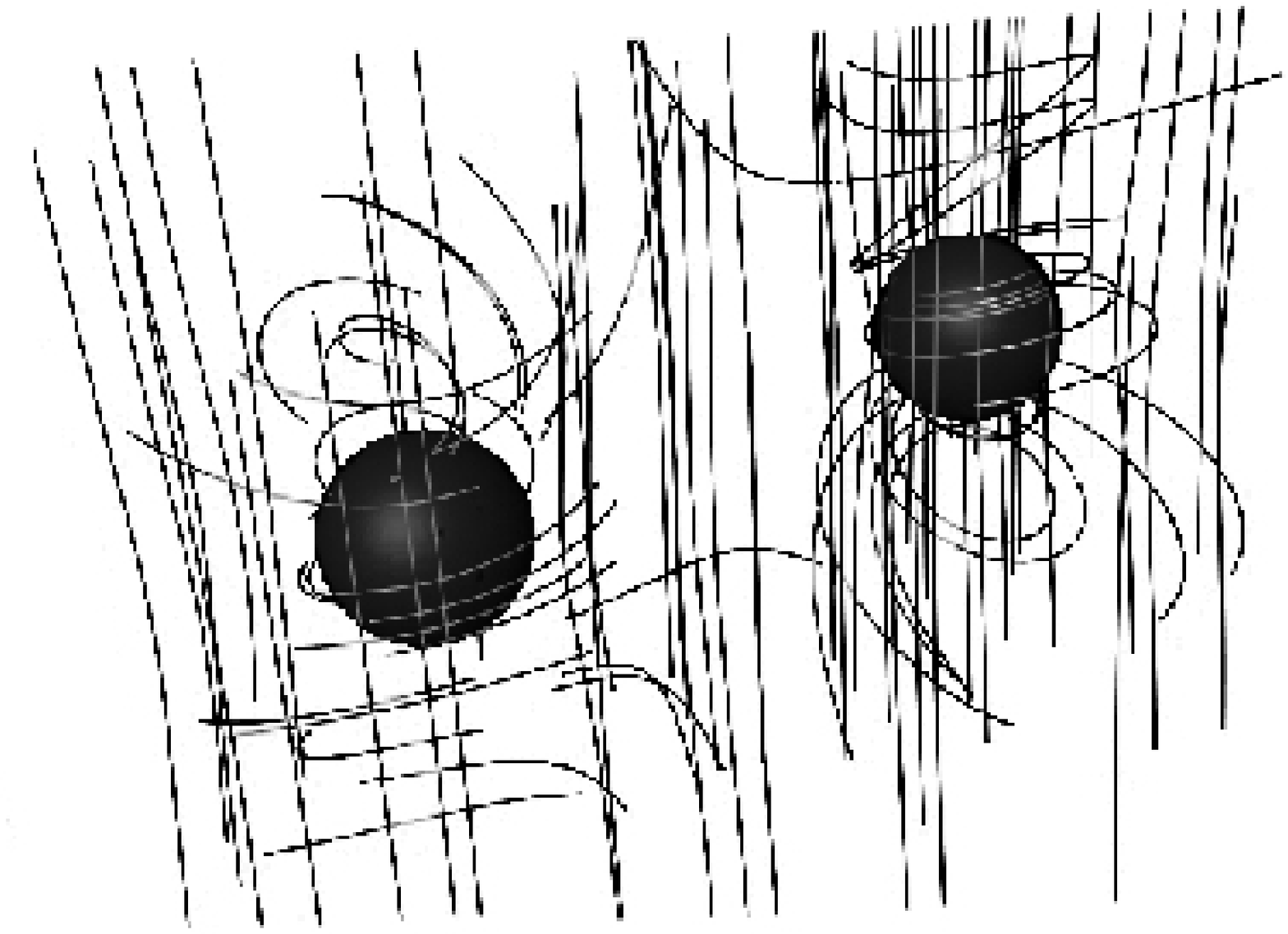,height=3.8cm,width=4.15cm}
\epsfig{file=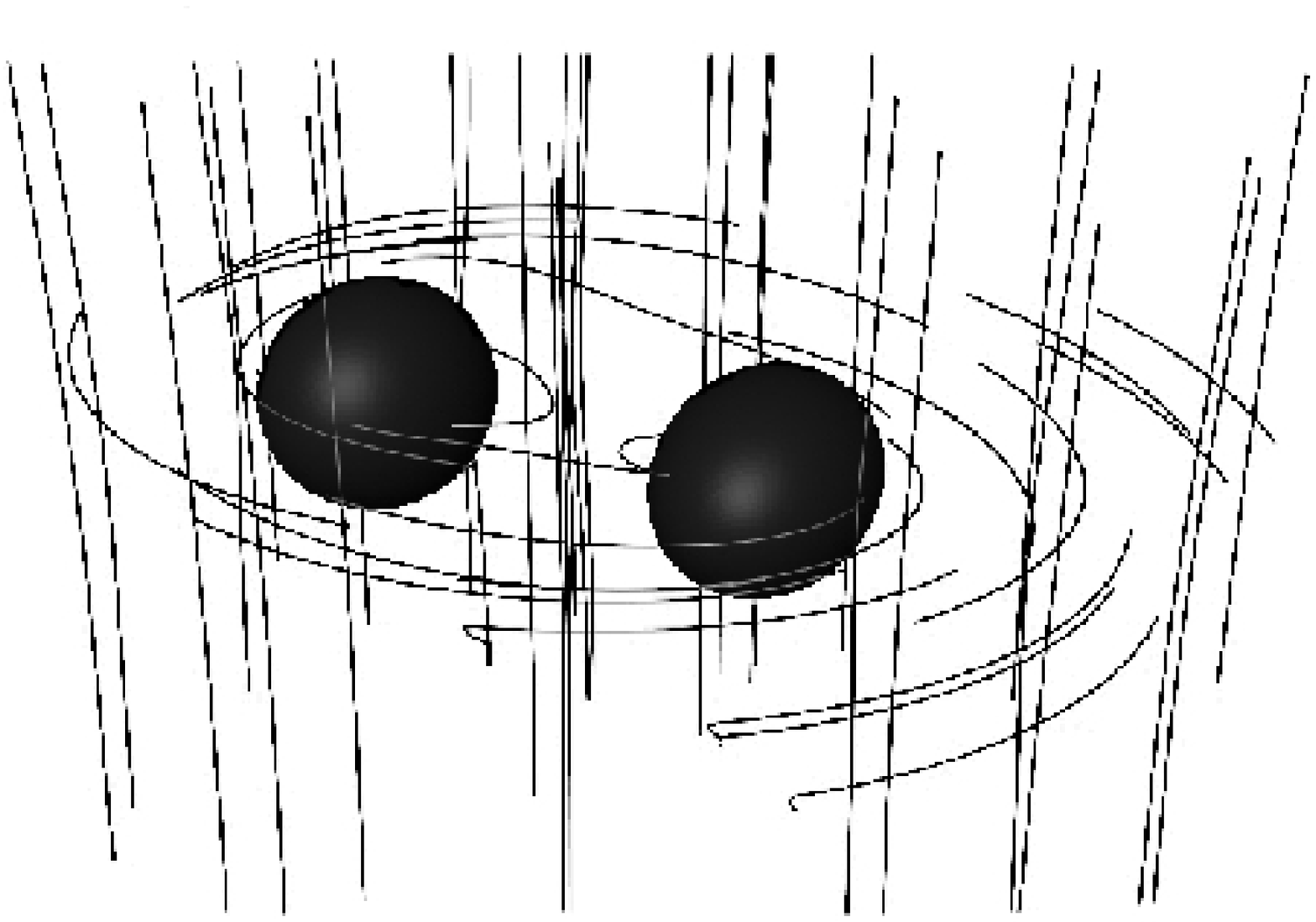,height=3.8cm,width=3.75cm}
\caption{Magnetic and electric field lines at $t \simeq -40M, -20M$ in 3D (the merger happens at $t=0$). The electric
field lines are twisted around the black hole, while the magnetic lines
are mostly aligned with the $z$-axis.}
\label{fig:binary_3dfield}
\end{center}
\end{figure}

Next, we consider a binary black hole system at a quasi-circular stage representing
a late inspiral. In the present work, we adopt initially non-spinning equal mass black holes 
to explore the dynamics of the system in a simple configuration.
The initial electromagnetic field is defined in exactly the same manner as in the single
black hole case. The evolution of this system, however, is significantly different
as the orbiting black holes stir the fields during the inspiral. Examples of
the field configurations during the evolution are provided in Fig.~\ref{fig:binary_3dfield}.

To analyze the influence of the binary's dynamics on the electromagnetic field we monitor two particular quantities. 
One is the electromagnetic energy density $E^2 + B^2$, while the
other is
%$T_{ab} n^a n^b$ defined in terms of the stress-energy tensor $T_{ab}$ and time-like normal vector $n^a$.
%The other is
the radial component of the Poynting vector, $S^r$ (radial from the origin). 
The former
is employed to get a sense of the energy variation in the EM field, while the latter is used to examine the induced
multipolar structure
and for comparisons to the gravitational wave signal computed from the Newman-Penrose  scalar,
$\Psi_4$. (We additionally
compute the analogous EM Newman-Penrose radiative quantity $\Phi_2$
% = F_{ab} n^a \bar m^b$ 
and confirm that the same features are present in it as in the Poynting vector).
Note that since the magnetic field is anchored in the disk, the magnetic field does not decay
with the distance from the binary, and this obscures the standard interpretations drawn from the Poynting field.
Nevertheless, the field does serve to elucidate the mode structure induced by the dynamics.  

\begin{figure}[h]
\begin{center}
\epsfig{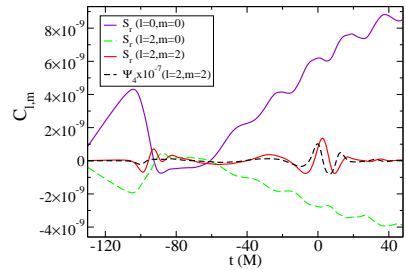}
\caption{The radial component of the Poynting vector for the binary black hole case. The structure
reveals it is dominated by (spherical harmonic)  ${}^{EM}c_{0,0}$ and ${}^{EM}c_{2,0}$ modes in the orbiting stage
though close to merger a ${}^{EM}c_{2,2}$ is induced which lags behind the 
corresponding gravitational wave (spin-2 spherical harmonic) ${}^{GW}c_{2,2}$. The 
transient behavior before $t\simeq -100$ is spurious and due to the initial data employed.}
\label{fig:multipoles}
\end{center}
\end{figure}

\begin{figure}[h]
\begin{center}
\epsfig{file=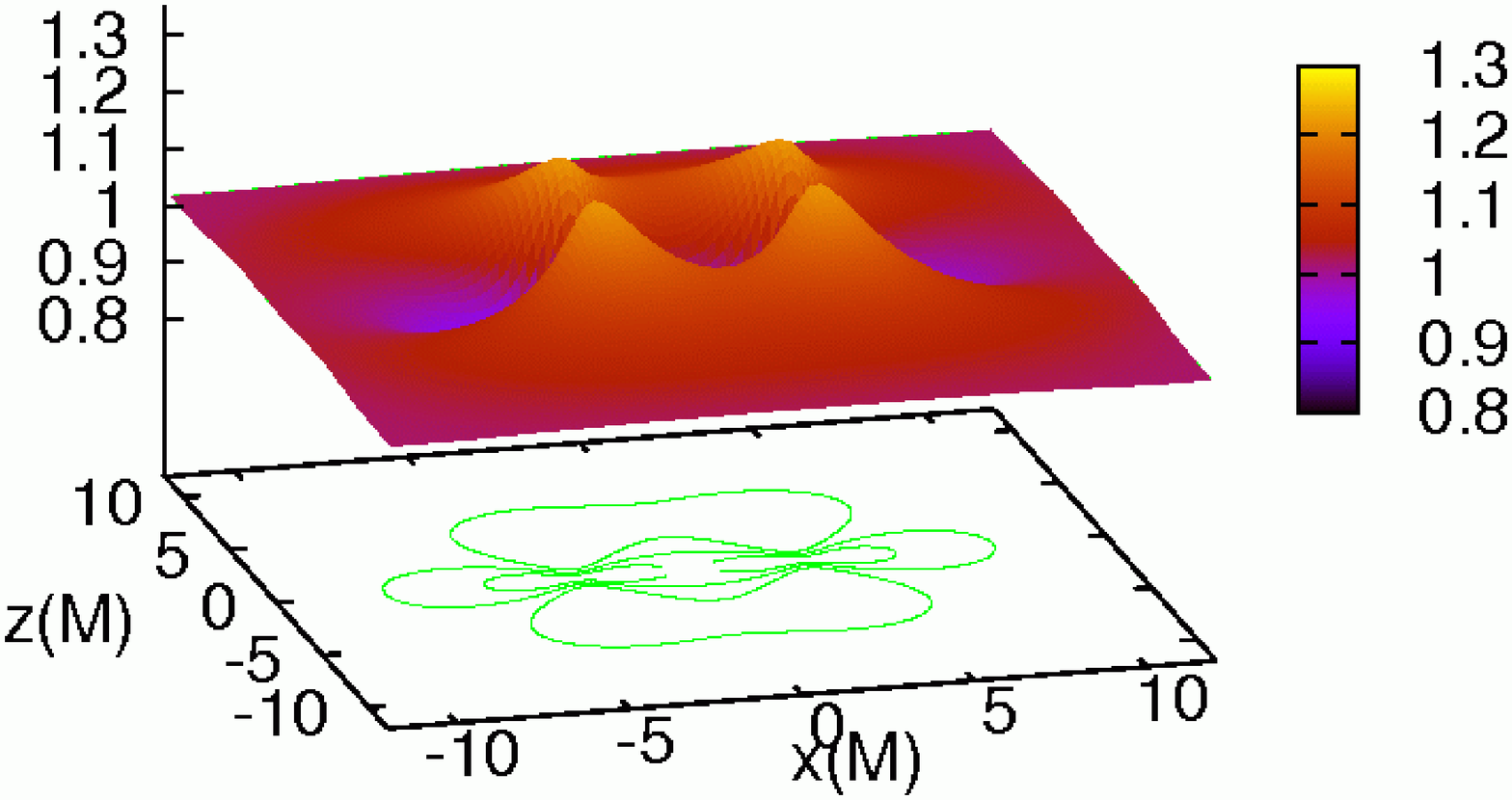,height=3.5cm,width=4.2cm} 
\epsfig{file=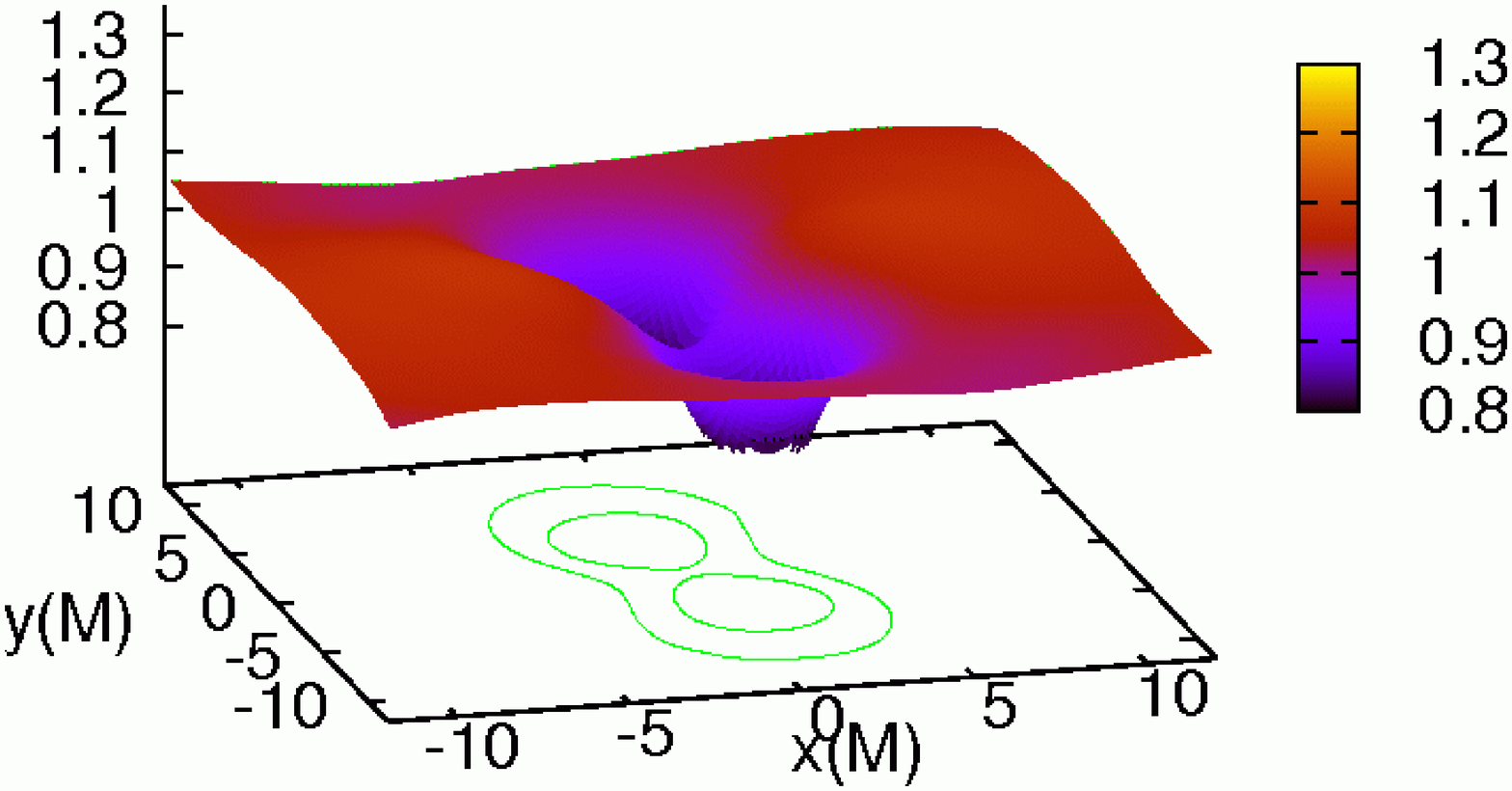,height=3.5cm,width=4.2cm} 
\epsfig{file=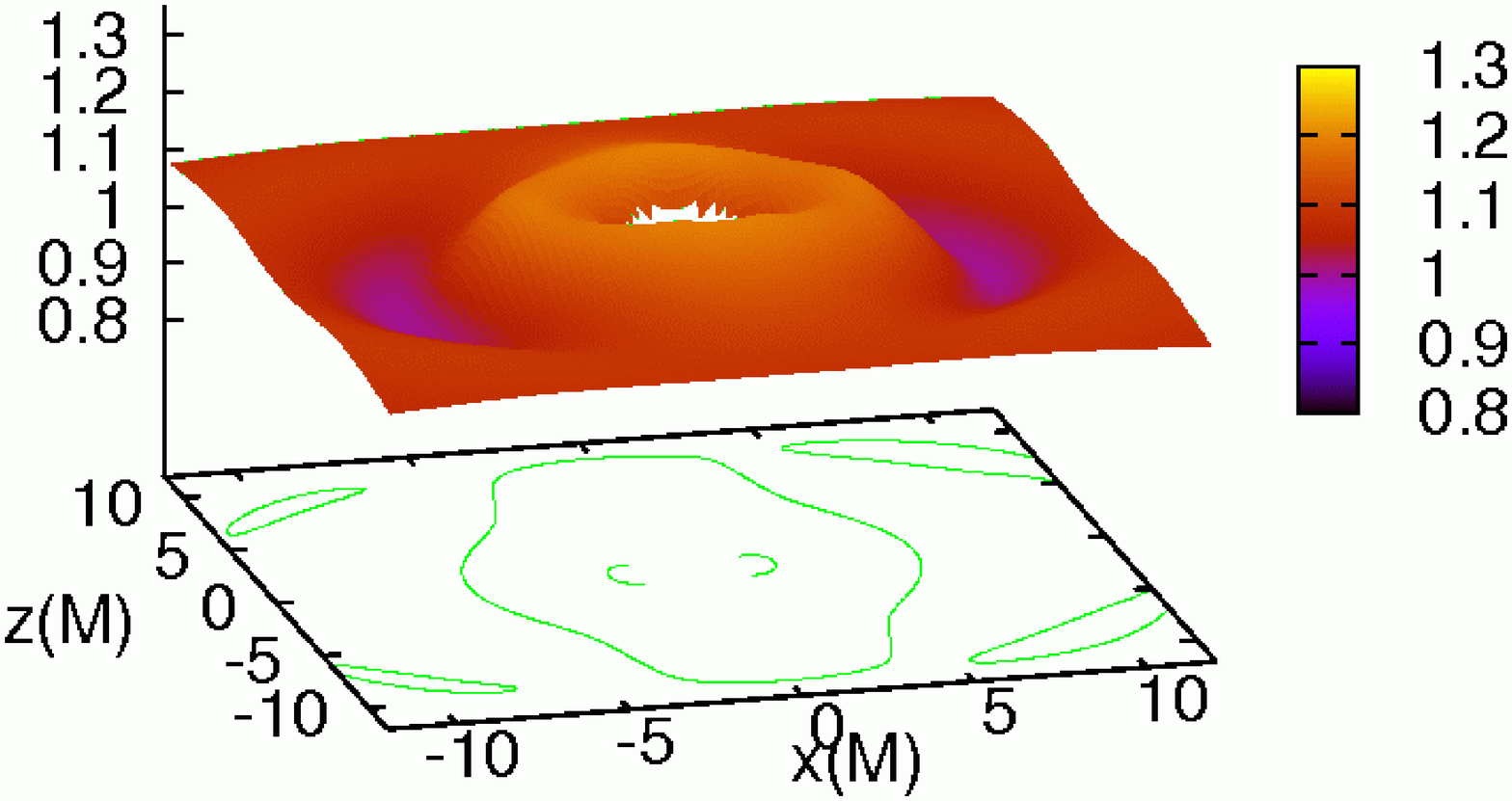,height=3.5cm,width=4.2cm} 
\epsfig{file=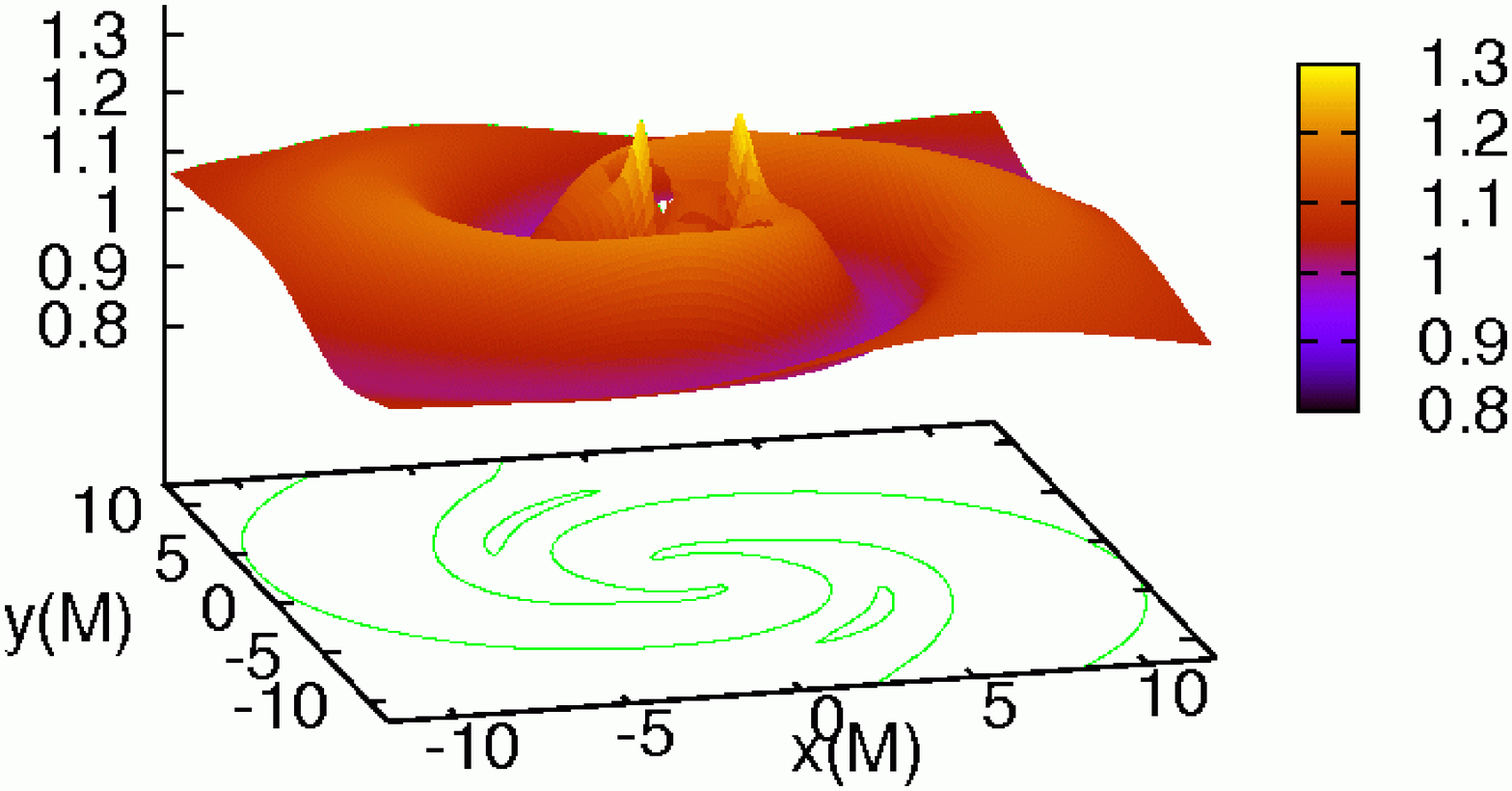,height=3.5cm,width=4.2cm} 
\epsfig{file=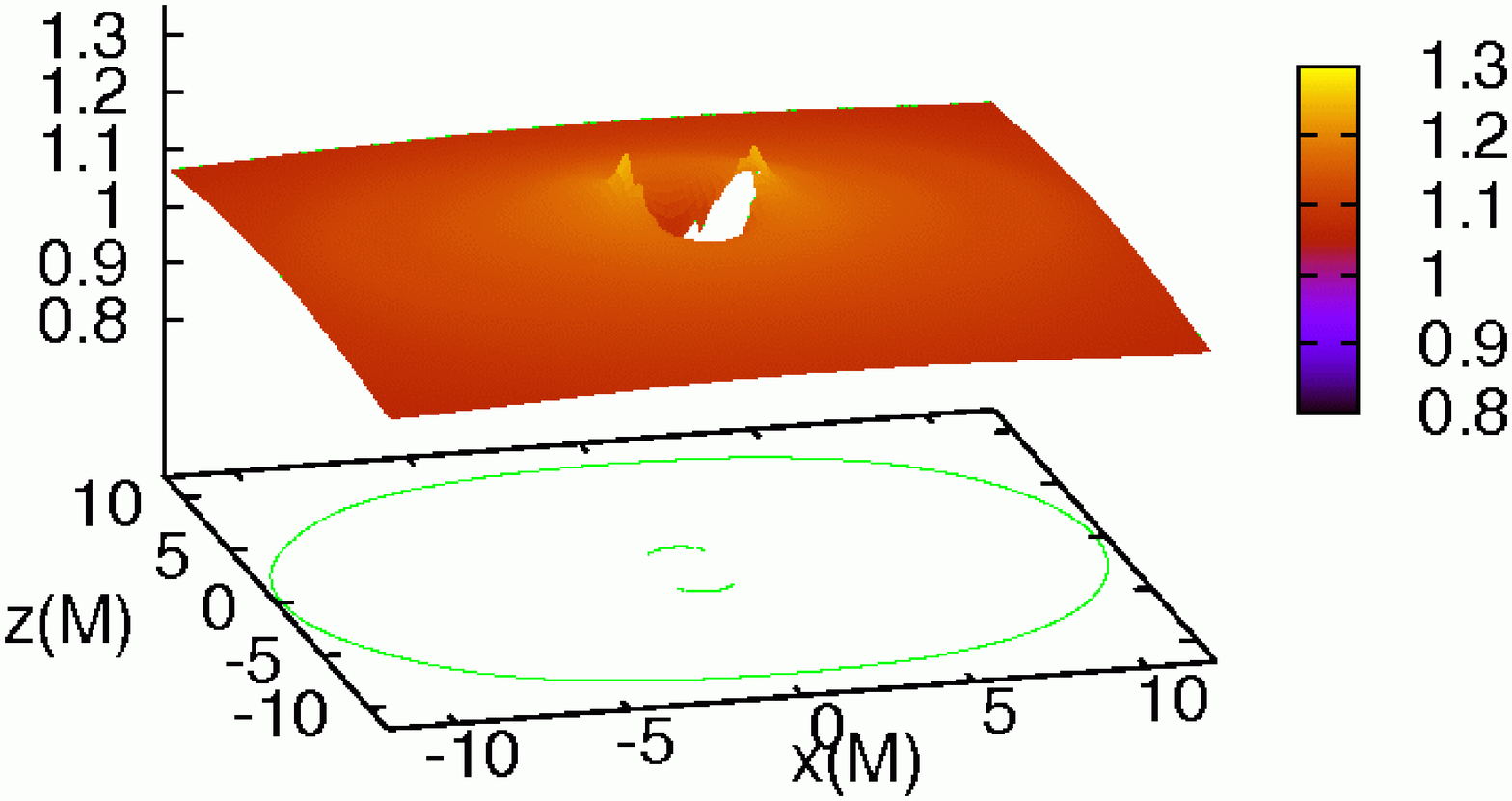,height=3.5cm,width=4.2cm} 
\epsfig{file=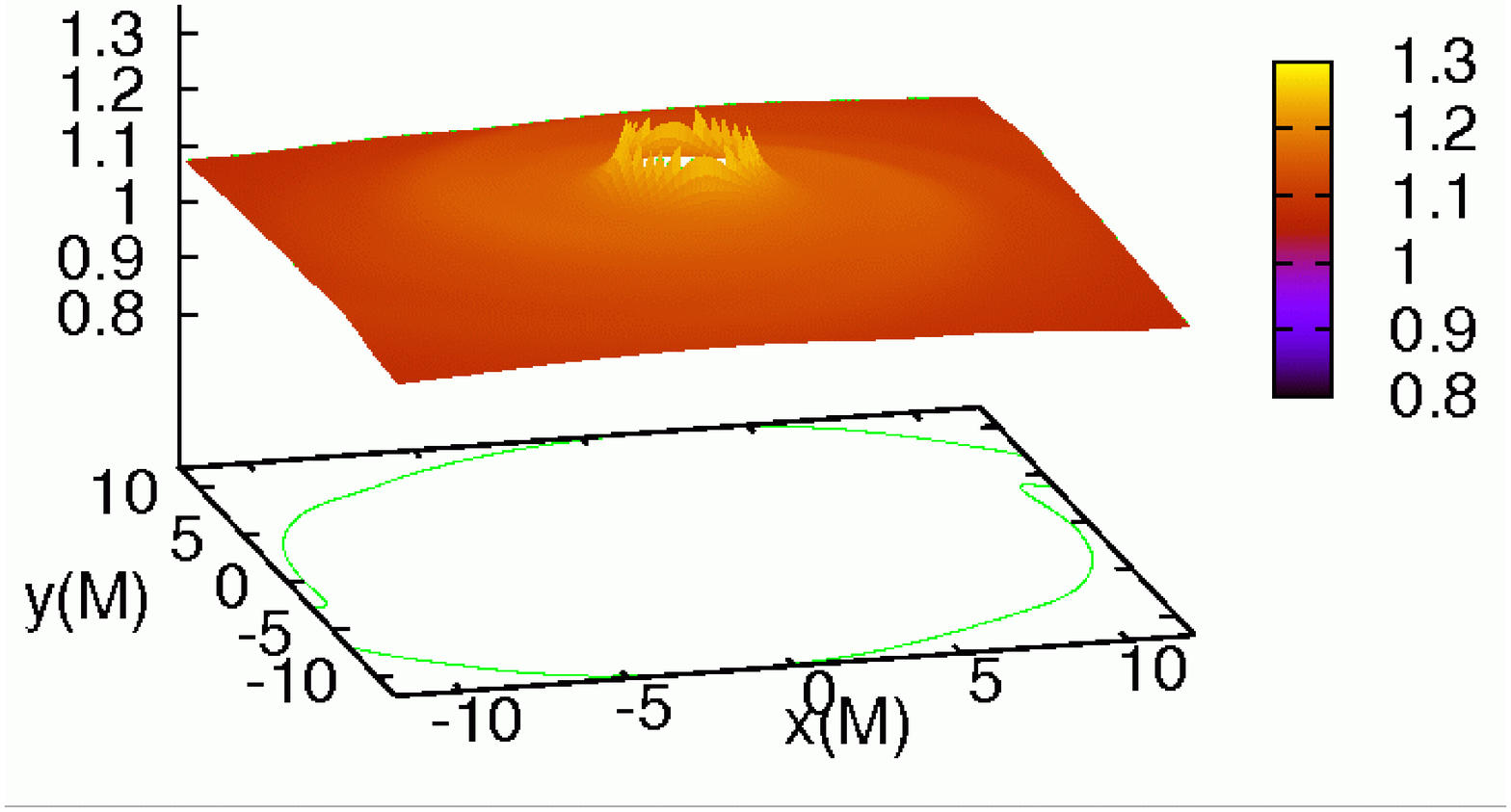,height=3.5cm,width=4.2cm} 
\epsfig{file=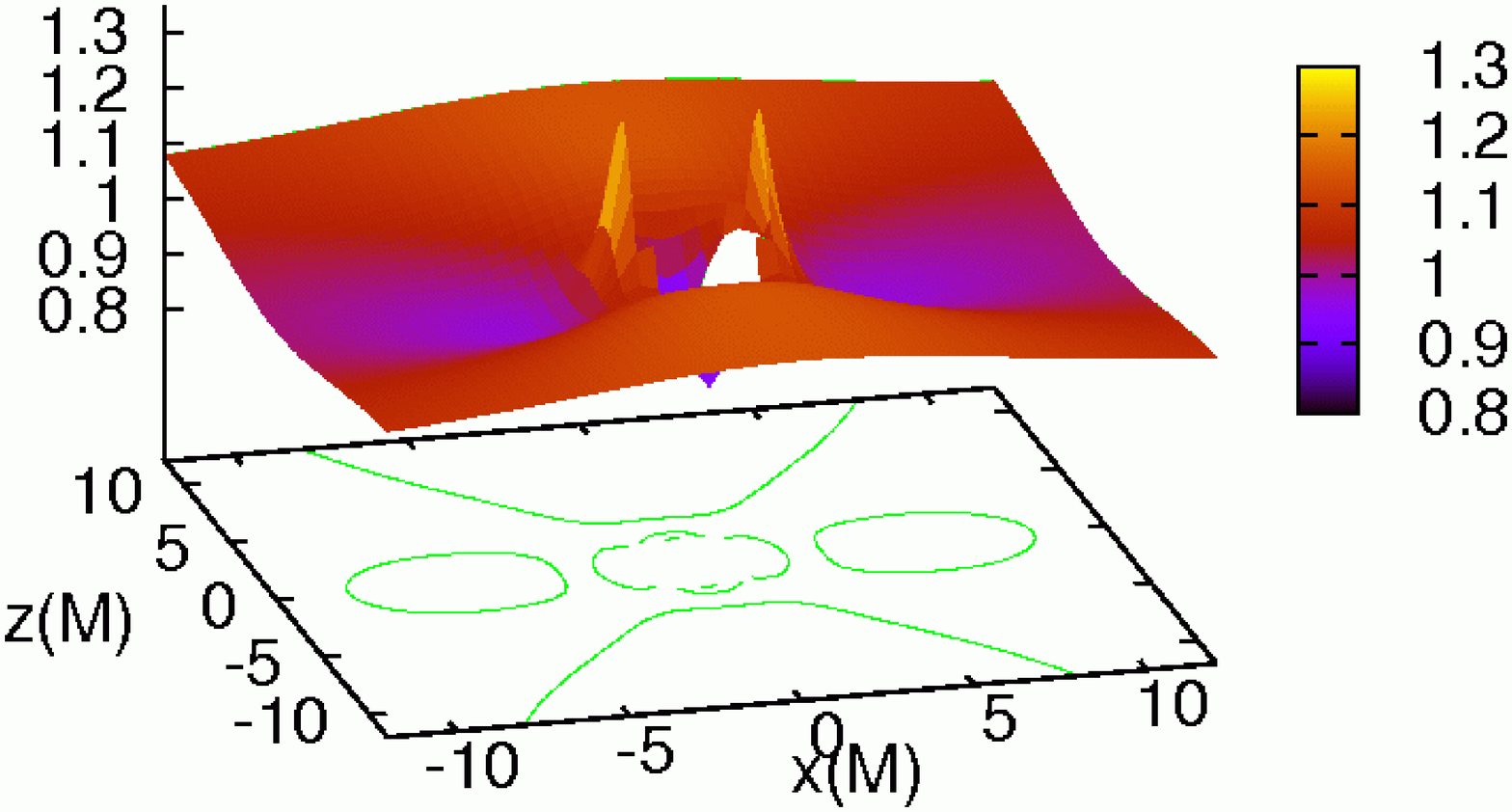,height=3.5cm,width=4.2cm} 
\epsfig{file=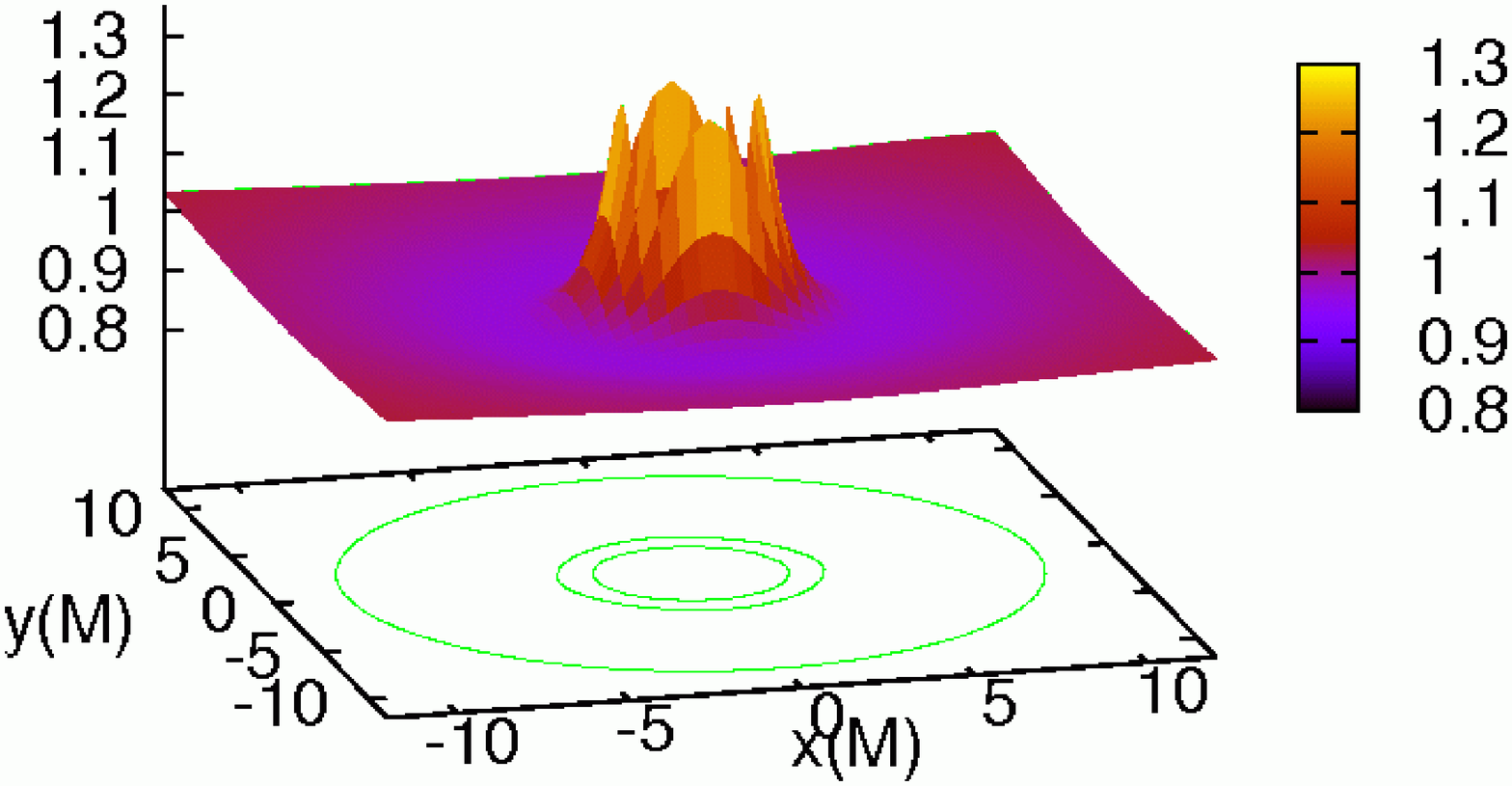,height=3.5cm,width=4.2cm} 
\caption{Normalized electromagnetic energy density (with respect to its value at the initial time in the absence
of black holes).  Left/Right columns corresponds to $y=0$ and $z=0$ plane while the first three rows correspond to 
times $t \simeq -70$, $5$ and $30M$ and the last to the single black hole case at $t=150M$ for the sake of comparison.
Normalized energy contour plots are shown with lines from $1.2$ to $0.8$ at intervals of $0.1$. Around the merger epoch, 
the energy computed for the binary case is $\simeq 30-40\%$ larger than in the single case.}
\label{fig:binary_B}
\end{center}
\end{figure}

An examination of these quantities reveals several distinct features during the evolution. 
After an initial transient signal,
and well
before the merger takes place, the electromagnetic fields display a pattern consistent with
those produced by electric dipoles, pointing along the orbital plane in the direction
orthogonal to their velocity, orbiting about each other.
This can be understood from the `membrane paradigm' point of view~\cite{1986bhmp.book.....T}, in which the 
stretched horizon is endowed with a surface density of electrical charge and a surface 
resistivity. The quasi-circular trajectories
of the black holes cause a charge separation in the direction perpendicular to both the
velocity and magnetic field, as in the Hall effect.
In particular, the evolution induces a toroidal electric field
 ($E_{\phi}$) with strength given by $B_{z} v_{\rm orbital} c^{-1}$. These fields 
would seed a Blandford-Znajek type mechanism and generate energy when interacting with a surrounding medium.  
Additionally, the electric and magnetic fields oscillate with a frequency equal to 
the dominant frequency of the gravitational radiation.
Since the Poynting flux is determined by the square of the electromagnetic fields,
its oscillation pattern has a frequency that is twice that of the dominant gravitational wave signal, and
can be described in terms of $l=0,m=0$ and $l=2,m=0$ modes. 
As the merger approaches, an $l=2,m=\pm 2$ mode is induced and the peak of EM radiated energy
lags behind the gravitational 
one by $t \simeq 2M$ ($ 1\times 10^{-5} (M/M_{\odot})$~s ) with the same frequency (see Fig.~\ref{fig:multipoles}). 
This lag time is likely related to the 
transition time from the binary black hole system to a single black hole, measured as
the `time of flight' from the ISCO to the light ring. Indeed estimates of the ISCO and the light ring (lr) 
for $a=0.67/M$ reveals $\Delta = r_{\hbox{\rm\scriptsize ISCO}} - r_{\mbox{\scriptsize lr}}\simeq 1.5M$.
Additionally, an enhancement of the electromagnetic field energy is clearly visible in the binary case, which displays
a $\simeq 30-40\%$ increase with respect to the single case, as illustrated in Fig.~\ref{fig:binary_B}. 
(We conjecture that in the presence of gas and black hole spins, this number could significanly increase). 
At late times, the system asymptotes to the same solution observed for the single black hole.

%\begin{figure}[h]
%\begin{center}
%\epsfig{file=figures/flux.eps,height=4.6cm,width=6.2cm}
%\caption{The flux of gravitational and electromagnetic energy for the binary black hole case. The maximum of the gravitational flux %happens at the merger, while the maximum of the
%electromagnetic flux (which has been reescaled for convenience) is shifted slightly after the merger.}
%\label{fig:flux}
%\end{center}
%\end{figure}

%\begin{figure}[h]
%\begin{center}
%\epsfig{file=figures/poynting_modes2.eps,height=4.6cm,width=6.2cm}
%\caption{The same as before but after filtering.} 
%\label{fig:multipoles}
%\end{center}
%\end{figure}

%\begin{figure}[h]
%\begin{center}
%\includegraphics[angle=-90,totalheight=0.2\textheight,width=0.42\textwidth]{hwaveforms.ps}
%\caption{and then some.
%\label{waveforms}
%}
%\end{center}
%\end{figure}

%%%%%%%%%%%%%%%%%%%%%%%%%%%%%%%%%%%%%%%%%%%%%%%%%%%%%%%%%%%%%%%%%%%%
%
%   S E C T I O N
%
%%%%%%%%%%%%%%%%%%%%%%%%%%%%%%%%%%%%%%%%%%%%%%%%%%%%%%%%%%%%%%%%%%%%
%\section{Conclusion}
\noindent{\bf{\em Final Comments:}} 
We have analyzed the behavior of electromagnetic fields influenced by
the dynamics of a binary black hole system. Our study illustrates several
interesting aspects of such systems that emit not only gravitational waves,
but
can also radiate electromagnetically when interacting with a plasma.

The EM fields have a clearly discernible pattern tied to the dynamics, making them
additional {\em tracers of the spacetime}, as these features would imprint particular characteristics
in processes producing observable EM signals. 
In particular, in the pre-merger stages, the black hole dynamics induce EM energy flux
oscillations with a period half that of the dominant GW signal
produced by the system, i.e., a fourth of the orbital period, and a gradual enhancement of 
the energy in the
electromagnetic field. During the merger, additional structure
is impressed on the field, yielding a distinct pattern that lags behind the gravitational
wave output by a time consistent with the characteristic time from the ISCO to the light ring of the
merged black hole~\footnote{Further studies with unequal masses or spins will help
test this conjecture.}. 

The distinct variability of the electromagnetic field as it
is dragged by the black holes suggests tantalizing prospects to
detect pre-merger electromagnetic signals from systems detectable in the gravitational
wave band. At a more speculative level, these combined signals could be 
exploited to study alternative
theories where photons and gravitons might propagate at different speeds 
(for a recent discussion of some possibilities see~\cite{Haiman:2008zy,Bloom:2009vx}).
A complete description of the problem requires the incorporation of gas and
radiation effects. However, the main qualitative features---driven by the orbiting
behavior of the black holes, whose inertia is many orders of magnitude above all 
else---would intuitively remain unaltered. The same reasoning also indicates the qualitative
features described above will be present in non-equal mass binaries. However, the
presence of spin could introduce further features, as energy could be extracted
from the individual black holes prior to merger.  
Ongoing~\cite{palenzuelaetal} and related work will help to test these conjectures and
shed further light on the problem. 
\\
%%%%%%%%%%%%%%%%%%%%%%%%%%%%%%%%%%%%%%%%%%%%%%%%%%%%%%%%%%%%%%%%%%%%
%
%   A C K N O W L E D G M E N T S
%
%%%%%%%%%%%%%%%%%%%%%%%%%%%%%%%%%%%%%%%%%%%%%%%%%%%%%%%%%%%%%%%%%%%%
\noindent{\bf{\em Acknowledgments:}}
We thank  
P. Chang, J. Frank, K. Menou, S. Phinney and S. Yoshida for stimulating  discussions. 
This work was supported by the NSF grants 
PHY-0803629 and PHY-0653375 to LSU,  PHY-0803615 and CCF-0832966
to BYU, and PHY-0803624 and CCF-0833090 to LIU.
Computations were done at BYU,
LONI, LSU, and TeraGrid.
LL acknowledges the Aspen Center for Physics for hospitality where
this work was started.
%%%%%%%%%%%%%%%%%%%%%%%%%%%%%%%%%%%%%%%%%%%%%%%%%%%%%%%%%%%%%%%%%%%%
%
%   B I B L I O G R A P H Y
%
%%%%%%%%%%%%%%%%%%%%%%%%%%%%%%%%%%%%%%%%%%%%%%%%%%%%%%%%%%%%%%%%%%%%
%\bibliography{./bhbhem_rev}

\begin{thebibliography}{33}
\expandafter\ifx\csname natexlab\endcsname\relax\def\natexlab#1{#1}\fi
\expandafter\ifx\csname bibnamefont\endcsname\relax
  \def\bibnamefont#1{#1}\fi
\expandafter\ifx\csname bibfnamefont\endcsname\relax
  \def\bibfnamefont#1{#1}\fi
\expandafter\ifx\csname citenamefont\endcsname\relax
  \def\citenamefont#1{#1}\fi
\expandafter\ifx\csname url\endcsname\relax
  \def\url#1{\texttt{#1}}\fi
\expandafter\ifx\csname urlprefix\endcsname\relax\def\urlprefix{URL }\fi
\providecommand{\bibinfo}[2]{#2}
\providecommand{\eprint}[2][]{\url{#2}}

\bibitem[{\citenamefont{Sylvestre}(2003)}]{Sylvestre:2003vc}
\bibinfo{author}{\bibfnamefont{J.}~\bibnamefont{Sylvestre}},
  \bibinfo{journal}{Astrophys. J.} \textbf{\bibinfo{volume}{591}},
  \bibinfo{pages}{1152} (\bibinfo{year}{2003}).

\bibitem[{\citenamefont{Stubbs}(2008)}]{Stubbs:2007mk}
\bibinfo{author}{\bibfnamefont{C.~W.} \bibnamefont{Stubbs}},
  \bibinfo{journal}{Class. Quant. Grav.} \textbf{\bibinfo{volume}{25}},
  \bibinfo{pages}{184033} (\bibinfo{year}{2008}).

\bibitem[{\citenamefont{Penrose}(1969)}]{Penrose:1969pc}
\bibinfo{author}{\bibfnamefont{R.}~\bibnamefont{Penrose}},
  \bibinfo{journal}{Riv. Nuovo Cim.} \textbf{\bibinfo{volume}{1}},
  \bibinfo{pages}{252} (\bibinfo{year}{1969}).

\bibitem[{\citenamefont{Blandford and Znajek}(1977)}]{Blandford:1977ds}
\bibinfo{author}{\bibfnamefont{R.~D.} \bibnamefont{Blandford}}
  \bibnamefont{and} \bibinfo{author}{\bibfnamefont{R.~L.}
  \bibnamefont{Znajek}}, \bibinfo{journal}{Mon. Not. Roy. Astron. Soc.}
  \textbf{\bibinfo{volume}{179}}, \bibinfo{pages}{433} (\bibinfo{year}{1977}).

\bibitem[{\citenamefont{{Begelman} et~al.}(1980)\citenamefont{{Begelman},
  {Blandford}, and {Rees}}}]{1980Natur.287..307B}
\bibinfo{author}{\bibfnamefont{M.~C.} \bibnamefont{{Begelman}}},
  \bibinfo{author}{\bibfnamefont{R.~D.} \bibnamefont{{Blandford}}},
  \bibnamefont{and} \bibinfo{author}{\bibfnamefont{M.~J.}
  \bibnamefont{{Rees}}}, \bibinfo{journal}{\nat}
  \textbf{\bibinfo{volume}{287}}, \bibinfo{pages}{307} (\bibinfo{year}{1980}).

\bibitem[{\citenamefont{Milosavljevic and
  Phinney}(2005)}]{Milosavljevic:2004cg}
\bibinfo{author}{\bibfnamefont{M.}~\bibnamefont{Milosavljevic}}
  \bibnamefont{and} \bibinfo{author}{\bibfnamefont{E.~S.}
  \bibnamefont{Phinney}}, \bibinfo{journal}{Astrophys. J.}
  \textbf{\bibinfo{volume}{622}}, \bibinfo{pages}{L93} (\bibinfo{year}{2005}).

\bibitem[{\citenamefont{Haiman et~al.}(2009)\citenamefont{Haiman, Kocsis, and
  Menou}}]{Haiman:2009te}
\bibinfo{author}{\bibfnamefont{Z.}~\bibnamefont{Haiman}},
  \bibinfo{author}{\bibfnamefont{B.}~\bibnamefont{Kocsis}}, \bibnamefont{and}
  \bibinfo{author}{\bibfnamefont{K.}~\bibnamefont{Menou}}
  (\bibinfo{year}{2009}), \eprint{0904.1383}.

\bibitem[{\citenamefont{{Artymowicz} and {Lubow}}(1996)}]{1996ApJ...467L..77A}
\bibinfo{author}{\bibfnamefont{P.}~\bibnamefont{{Artymowicz}}}
  \bibnamefont{and} \bibinfo{author}{\bibfnamefont{S.~H.}
  \bibnamefont{{Lubow}}}, \bibinfo{journal}{\apjl}
  \textbf{\bibinfo{volume}{467}}, \bibinfo{pages}{L77+} (\bibinfo{year}{1996}).

\bibitem[{\citenamefont{{MacFadyen} and
  {Milosavljevi{\'c}}}(2008)}]{2008ApJ...672...83M}
\bibinfo{author}{\bibfnamefont{A.~I.} \bibnamefont{{MacFadyen}}}
  \bibnamefont{and}
  \bibinfo{author}{\bibfnamefont{M.}~\bibnamefont{{Milosavljevi{\'c}}}},
  \bibinfo{journal}{\apj} \textbf{\bibinfo{volume}{672}}, \bibinfo{pages}{83}
  (\bibinfo{year}{2008}).

\bibitem[{\citenamefont{{Cuadra} et~al.}(2009)}]{2009MNRAS.393.1423C}
\bibinfo{author}{\bibfnamefont{J.}~\bibnamefont{{Cuadra}}}
  \bibnamefont{et~al.}, \bibinfo{journal}{\mnras}
  \textbf{\bibinfo{volume}{393}}, \bibinfo{pages}{1423} (\bibinfo{year}{2009}).

\bibitem[{\citenamefont{{Armitage} and
  {Natarajan}}(2002)}]{2002ApJ...567L...9A}
\bibinfo{author}{\bibfnamefont{P.~J.} \bibnamefont{{Armitage}}}
  \bibnamefont{and}
  \bibinfo{author}{\bibfnamefont{P.}~\bibnamefont{{Natarajan}}},
  \bibinfo{journal}{\apjl} \textbf{\bibinfo{volume}{567}}, \bibinfo{pages}{L9}
  (\bibinfo{year}{2002}).

\bibitem[{\citenamefont{{Liu} et~al.}(2003)\citenamefont{{Liu}, {Wu}, and
  {Cao}}}]{2003MNRAS.340..411L}
\bibinfo{author}{\bibfnamefont{F.~K.} \bibnamefont{{Liu}}},
  \bibinfo{author}{\bibfnamefont{X.-B.} \bibnamefont{{Wu}}}, \bibnamefont{and}
  \bibinfo{author}{\bibfnamefont{S.~L.} \bibnamefont{{Cao}}},
  \bibinfo{journal}{\mnras} \textbf{\bibinfo{volume}{340}},
  \bibinfo{pages}{411} (\bibinfo{year}{2003}).

\bibitem[{\citenamefont{Palenzuela et~al.}(2007)}]{Palenzuela:2006wp}
\bibinfo{author}{\bibfnamefont{C.}~\bibnamefont{Palenzuela}}
  \bibnamefont{et~al.}, \bibinfo{journal}{Phys. Rev.}
  \textbf{\bibinfo{volume}{D75}}, \bibinfo{pages}{064005}
  (\bibinfo{year}{2007}).

\bibitem[{\citenamefont{{Komissarov}}(2007)}]{2007MNRAS.382..995K}
\bibinfo{author}{\bibfnamefont{S.~S.} \bibnamefont{{Komissarov}}},
  \bibinfo{journal}{\mnras} \textbf{\bibinfo{volume}{382}},
  \bibinfo{pages}{995} (\bibinfo{year}{2007}).

\bibitem[{\citenamefont{Gundlach et~al.}(2005)}]{Gundlach:2005eh}
\bibinfo{author}{\bibfnamefont{C.}~\bibnamefont{Gundlach}}
  \bibnamefont{et~al.}, \bibinfo{journal}{Class. Quant. Grav.}
  \textbf{\bibinfo{volume}{22}}, \bibinfo{pages}{3767} (\bibinfo{year}{2005}).

\bibitem[{\citenamefont{Dedner et~al.}(2002)}]{Dedner2002}
\bibinfo{author}{\bibfnamefont{A.}~\bibnamefont{Dedner}} \bibnamefont{et~al.},
  \bibinfo{journal}{J. Comput. Phys.} \textbf{\bibinfo{volume}{175}},
  \bibinfo{pages}{645} (\bibinfo{year}{2002}).

\bibitem[{\citenamefont{Rinne et~al.}(2007)\citenamefont{Rinne, Lindblom, and
  Scheel}}]{RLS07}
\bibinfo{author}{\bibfnamefont{O.}~\bibnamefont{Rinne}},
  \bibinfo{author}{\bibfnamefont{L.}~\bibnamefont{Lindblom}}, \bibnamefont{and}
  \bibinfo{author}{\bibfnamefont{M.~A.} \bibnamefont{Scheel}},
  \bibinfo{journal}{Class. Quant. Grav.} \textbf{\bibinfo{volume}{24}},
  \bibinfo{pages}{4054} (\bibinfo{year}{2007}).

\bibitem[{had()}]{had_webpage}
\bibinfo{note}{Http://www.had.liu.edu/}.

\bibitem[{\citenamefont{Liebling}(2002)}]{Liebling}
\bibinfo{author}{\bibfnamefont{S.~L.} \bibnamefont{Liebling}},
  \bibinfo{journal}{Phys. Rev.} \textbf{\bibinfo{volume}{D66}},
  \bibinfo{pages}{041703} (\bibinfo{year}{2002}).

\bibitem[{\citenamefont{Lehner et~al.}(2006)\citenamefont{Lehner, Liebling, and
  Reula}}]{Lehner:2005vc}
\bibinfo{author}{\bibfnamefont{L.}~\bibnamefont{Lehner}},
  \bibinfo{author}{\bibfnamefont{S.~L.} \bibnamefont{Liebling}},
  \bibnamefont{and} \bibinfo{author}{\bibfnamefont{O.}~\bibnamefont{Reula}},
  \bibinfo{journal}{Class. Quant. Grav.} \textbf{\bibinfo{volume}{23}},
  \bibinfo{pages}{S421} (\bibinfo{year}{2006}).

\bibitem[{\citenamefont{Anderson et~al.}(2008)}]{binaryNS}
\bibinfo{author}{\bibfnamefont{M.}~\bibnamefont{Anderson}}
  \bibnamefont{et~al.}, \bibinfo{journal}{Phys. Rev.}
  \textbf{\bibinfo{volume}{D77}}, \bibinfo{pages}{024006}
  (\bibinfo{year}{2008}).

\bibitem[{lor()}]{lorene_webpage}
\bibinfo{note}{Http://www.lorene.obspm.fr/}.

\bibitem[{\citenamefont{Buonanno et~al.}(2008)\citenamefont{Buonanno, Kidder,
  and Lehner}}]{Buonanno:2007sv}
\bibinfo{author}{\bibfnamefont{A.}~\bibnamefont{Buonanno}},
  \bibinfo{author}{\bibfnamefont{L.~E.} \bibnamefont{Kidder}},
  \bibnamefont{and} \bibinfo{author}{\bibfnamefont{L.}~\bibnamefont{Lehner}},
  \bibinfo{journal}{Phys. Rev.} \textbf{\bibinfo{volume}{D77}},
  \bibinfo{pages}{026004} (\bibinfo{year}{2008}).

\bibitem[{\citenamefont{Jackson}(1975)}]{Jackson1975}
\bibinfo{author}{\bibfnamefont{J.~D.} \bibnamefont{Jackson}},
  \emph{\bibinfo{title}{Classical Electrodynamics}} (\bibinfo{publisher}{John
  Wiley \& Sons}, \bibinfo{address}{New York}, \bibinfo{year}{1975}).

\bibitem[{\citenamefont{{Massi} and {Kaufman}}(2008)}]{2008A&A...477....1M}
\bibinfo{author}{\bibfnamefont{M.}~\bibnamefont{{Massi}}} \bibnamefont{and}
  \bibinfo{author}{\bibfnamefont{M.}~\bibnamefont{{Kaufman}}},
  \bibinfo{journal}{\aap} \textbf{\bibinfo{volume}{477}}, \bibinfo{pages}{1}
  (\bibinfo{year}{2008}).

\bibitem[{\citenamefont{King et~al.}(1975)\citenamefont{King, Lasota, and
  Kundt}}]{KiLaKu75}
\bibinfo{author}{\bibfnamefont{A.~R.} \bibnamefont{King}},
  \bibinfo{author}{\bibfnamefont{J.~P.} \bibnamefont{Lasota}},
  \bibnamefont{and} \bibinfo{author}{\bibfnamefont{W.}~\bibnamefont{Kundt}},
  \bibinfo{journal}{\prd} \textbf{\bibinfo{volume}{12}}, \bibinfo{pages}{3037}
  (\bibinfo{year}{1975}).

\bibitem[{\citenamefont{Komissarov and McKinney}(2007)}]{Komissarov:2007rc}
\bibinfo{author}{\bibfnamefont{S.~S.} \bibnamefont{Komissarov}}
  \bibnamefont{and} \bibinfo{author}{\bibfnamefont{J.~C.}
  \bibnamefont{McKinney}}, \bibinfo{journal}{Mon. Not. Roy. Astron. Soc. Lett.}
  \textbf{\bibinfo{volume}{377}}, \bibinfo{pages}{L49} (\bibinfo{year}{2007}).

\bibitem[{\citenamefont{{Wald}}(1974)}]{1974PhRvD..10.1680W}
\bibinfo{author}{\bibfnamefont{R.~M.} \bibnamefont{{Wald}}},
  \bibinfo{journal}{\prd} \textbf{\bibinfo{volume}{10}}, \bibinfo{pages}{1680}
  (\bibinfo{year}{1974}).

\bibitem[{\citenamefont{{Thorne} et~al.}(1986)\citenamefont{{Thorne}, {Price},
  and {MacDonald}}}]{1986bhmp.book.....T}
\bibinfo{author}{\bibfnamefont{K.~S.} \bibnamefont{{Thorne}}},
  \bibinfo{author}{\bibfnamefont{R.~H.} \bibnamefont{{Price}}},
  \bibnamefont{and} \bibinfo{author}{\bibfnamefont{D.~A.}
  \bibnamefont{{MacDonald}}}, \emph{\bibinfo{title}{{Black holes: The membrane
  paradigm}}} (\bibinfo{year}{1986}).

\bibitem[{\citenamefont{Haiman et~al.}(2008)}]{Haiman:2008zy}
\bibinfo{author}{\bibfnamefont{Z.}~\bibnamefont{Haiman}} \bibnamefont{et~al.}
  (\bibinfo{year}{2008}), \eprint{0811.1920}.

\bibitem[{\citenamefont{Bloom et~al.}(2009)}]{Bloom:2009vx}
\bibinfo{author}{\bibfnamefont{J.~S.} \bibnamefont{Bloom}} \bibnamefont{et~al.}
  (\bibinfo{year}{2009}), \eprint{0902.1527}.

\bibitem[{\citenamefont{Palenzuela et~al.}(2009)}]{palenzuelaetal}
\bibinfo{author}{\bibfnamefont{C.}~\bibnamefont{Palenzuela}}
  \bibnamefont{et~al.}, \bibinfo{journal}{in preparation}
  (\bibinfo{year}{2009}).

\bibitem[{\citenamefont{Aylott et~al.}(2009)}]{Aylott:2009ya}
\bibinfo{author}{\bibfnamefont{B.}~\bibnamefont{Aylott}} \bibnamefont{et~al.}
  (\bibinfo{year}{2009}), \eprint{0901.4399}.

\end{thebibliography}
%\bibliographystyle{apsrev}
%\begin{thebibliography}{32}

%\end{thebibliography}

%%%%%%%%%%%%%%%%%%%%%%%%%%%%%%%%%%%%%%%%%%%%%%%%%%%%%%%%%%%%%%%%%%%%
%
%   E N D   D O C U M E N T
%
%%%%%%%%%%%%%%%%%%%%%%%%%%%%%%%%%%%%%%%%%%%%%%%%%%%%%%%%%%%%%%%%%%%%
\end{document}